\documentclass[conference,a4paper]{IEEEtran}
\usepackage[dvips]{graphicx}
\usepackage[utf8]{inputenc}
\usepackage[T1]{fontenc}
\usepackage{diagbox}

\usepackage[left=14mm, right=14mm, top=20mm, bottom=30mm]{geometry}  %  for two-column form
\usepackage{amsmath,amssymb,amsfonts,amsthm,mathrsfs}
\usepackage{epsfig,epsf,subfigure,graphicx,graphics}
\usepackage{url,enumerate}
\usepackage{color}
\usepackage{cite}
\usepackage{dblfloatfix,fixltx2e}
\usepackage{CJK,tikz}
\usepackage{multirow}
\usepackage{algorithm,algorithmic}
\begin{document}

\title{Reducing Cubic Metric of Circularly Pulse-Shaped OFDM Signals Through Constellation Shaping Optimization With Performance Constraints}
%A New Waveform and Its Transceiver Optimization
%\bstctlcite{IEEEexample:BSTcontrol}

\author{
\IEEEauthorblockN{Yenming Huang, Rueibin Yang, and Borching Su}
\IEEEauthorblockA{Graduate Institute of Communication Engineering, National Taiwan University, Taipei 10617, Taiwan\\
Email: \{d01942015, r06942058, borching\}@ntu.edu.tw}
}

%\markboth{IEEE Access Special Section on New Waveform Design and Air-Interface for Future Heterogeneous Network towards 5G}{Huang \MakeLowercase{\textit{et al.}}: Circularly Pulse-Shaped Precoding for OFDM: A New Waveform and Its Optimization Design for 5G New Radio}

\maketitle

%==============================================================%
%================== Abstract and Key Words ====================%
%==============================================================%
\begin{abstract}
Circularly pulse-shaped orthogonal frequency division multiplexing (CPS-OFDM) is one of the most promising 5G waveforms that addresses two physical layer signal requirements of low out-of-subband emission (OSBE) and low peak-to-average power ratio (PAPR) with flexibility in parameter adaptation.
In this paper, a constellation shaping optimization method is proposed to further reduce the cubic metric (CM) of CPS-OFDM signals for the case that demands rather high power amplifier (PA) efficiency at the transmitter.
Simulation results demonstrate the superiority of the proposed scheme in CM reduction, and the corresponding benefits of spectral regrowth mitigation and spectral efficiency improvement.
\end{abstract}
\begin{IEEEkeywords}
5G, New Radio (NR), new waveform, circularly pulse-shaped OFDM (CPS-OFDM), cubic metric (CM), constellation shaping, convex optimization, transceiver design.
\end{IEEEkeywords}

%==============================================================%
%========================= Contents ===========================%
%==============================================================%

%==============================================================%
\section{Introduction}
\label{Sec:Introduction}
%==============================================================%
The fifth generation wireless systems (5G), named as New Radio (NR), are envisioned to support various innovative service-oriented applications in manifold usage scenarios \cite{3GPPTR38913}, e.g., machine type device-to-device communications \cite{Sexton2018_WD}, grant-free asynchronous transmissions \cite{Wunder2014_WD}, and coexistence of heterogeneous spectrum access with different numerologies \cite{Huang2016_HYM}, etc. The physical layer signal formats of 5G NR, on basis of orthogonal frequency division multiplexing (OFDM) \cite{3GPPTR38912}, additionally demand low out-of-subband emission (OSBE), low peak-to-average power ratio (PAPR), flexibility in parameter adaptation, backward and forward compatibility \cite{Lien2017_HYM,Ankarali2017_WD,Zaidi2016_WD}.
Circularly pulse-shaped OFDM (CPS-OFDM) is one of the most promising waveforms that addresses these requirements simultaneously with spectral efficiency improvement \cite{Huang2018_HYM}. In contrary to a number of existing waveform schemes relying on windowing or filtering techniques \cite{Zhang2016_WD}, CPS-OFDM exploits a subband-wise precoder characterized by a prototype vector and its circularly time-frequency shifted versions without imposing guard interval (GI) burden \cite{Huang2018_HYM}. Besides, CPS-OFDM can be regarded as a generalized discrete Fourier transform spread OFDM (DFT-S-OFDM) implemented with linearithmic-order complexity \cite{Huang2018_HYM}. As CPS-OFDM has been demonstrated to bring such attractive advantages into upcoming 5G NR, it is worthy to study some auxiliary techniques that can further enhance the CPS-OFDM system performance in certain cases.

Considering the case that demands high power amplifier (PA) efficiency at the transmitter (e.g., a smart factory with intercommunicating machinery \cite{Sexton2018_WD}), this paper intends to further reduce the cubic metric (CM) of CPS-OFDM signals through a constellation shaping technique. CM is known as a more accurate indicator to quantify envelope fluctuations than PAPR \cite{Motorola040642_3GPPTDOC,Motorola060023_3GPPTDOC,Behravan2006_WDIS_PAE,Ni2017_WDIS_PAE_CM,Zhu2014_WDIS_PAE_CM,Rahmatallah2013_WDIS_PAE_PAPR}.
The reason behind this is that CM closely relates to the third-order intermodulation product of PA nonlinearity, which dominates the distortion of the transmitted signals \cite{Motorola040642_3GPPTDOC,Rahmatallah2013_WDIS_PAE_PAPR}.
The key idea of constellation shaping is to introduce offset values to input quadrature amplitude modulation (QAM) symbols so that CM (or PAPR) can be greatly reduced \cite{Rahmatallah2013_WDIS_PAE_PAPR,Zhu2014_WDIS_PAE_CM,Aggarwal2006_WDIS_PAE_PAPR}.
The use of this approach is usually accompanied with an error vector magnitude (EVM) constraint that confines the resulting detection performance degradation within an acceptable range.
Moreover, the inherent property of low OSBE of CPS-OFDM must be carefully preserved to facilitate asynchronous transmissions and mixed numerologies for lack of frequency domain user orthogonality in the case.

In this paper, we propose a constellation shaping optimization method for CPS-OFDM according to the aforementioned design aspects. 
Specifically, the proposed optimization problem is formulated to minimize the CM subject to two constraints on the EVM and the OSBE. 
The optimized offset values imposed on the transmitted QAM symbols are transparent to the receiver, i.e., no additional side information about applying constellation shaping to the transmitter is required by the receiver. 
Although the real-time optimization may encounter some challenges such as high computational complexity and algorithm reliability, we believe that the advances in electronic technology can afford high computing power for signal processing in devices in the near future \cite{Aggarwal2006_WDIS_PAE_PAPR,Mattingley2010_OPT}.

The rest of this paper is organized as follows.
In Section \ref{Sec:SystemMetric}, a CPS-OFDM transceiver system model and its design problem are described.
In Section \ref{Sec:ProposedMethod}, the proposed constellation shaping optimization method for CPS-OFDM is introduced. 
In Section \ref{Sec:SimResult}, simulation results show the performance gains of applying the proposed scheme to 5G NR. 
Finally, the contributions of this work are summarized in Section \ref{Sec:Conclusion}.

\begin{figure*}[t]
\centering \centerline{
\includegraphics[width=1.0\textwidth,clip]{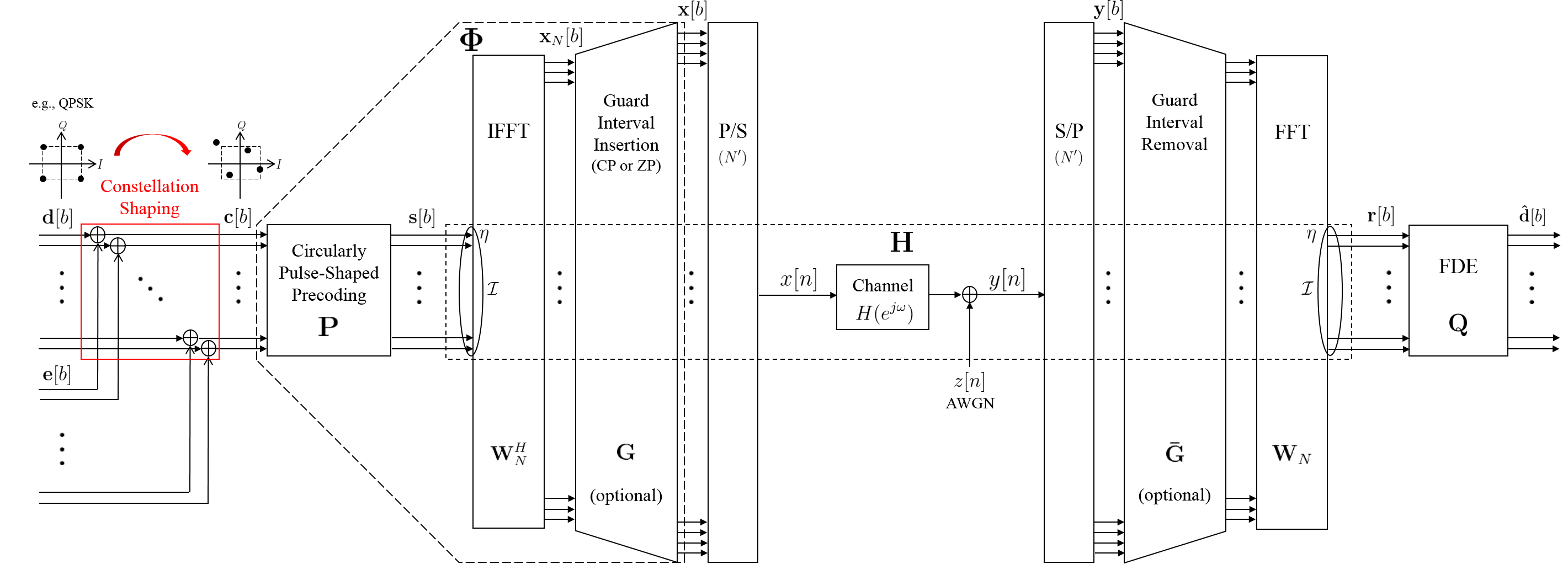}}
\caption{CPS-OFDM baseband transceiver system model with constellation shaping.}
\label{TxRx_ConstelShapedCPSOFDM}
\end{figure*} 

{\it Notations}:
Boldfaced lower case letters such as $\mathbf{x}$ represent column vectors, boldfaced upper case letters such as $\mathbf{X}$ represent matrices, and italic letters such as $X$ represent scalars.
Superscripts as in $\mathbf{X}^{H}$ and $\mathbf{X}^{-1}$ denote the transpose-conjugate and inverse operators, respectively.
Calligraphic upper case letters such as $\mathcal{I}$ represent sets of discrete indices or continuous intervals.
The submatrix of the matrix $\mathbf{X}$ formed by the column vectors with the ordered indices given in $\mathcal{I}$ is denoted by $\left[ \mathbf{X} \right] _{\mathcal{I}}$.
Similarly, the subvector of the column vector $\mathbf{x}$ is denoted as $\left[ \mathbf{x} \right]_{\mathcal{I}}$. 
Let $\mathbf{0}_{N\times M}$ and $\mathbf{I}_{N}$ be the $N\times M$ zero matrix and the $N\times N$ identity matrix, respectively. 
The $(N\times M)$-dimensional complex matrix space is expressed as $\mathbb{C}^{N\times M}$.
Operators $\left \langle \cdot \right \rangle _{S}$, $\left | \cdot  \right |$, $\left \| \cdot \right \|_{2}$, $\left \| \cdot \right \|_{6}$, and $\Re \{ \cdot \}$ denote modulo $S$, modulus of a complex scalar, Euclidean norm, 6-norm, and real part of a complex number, respectively. 
The decibel (dB) representation of a real number $C$ is shown as $C|_{\textrm{dB}}$.
Throughout the paper we adopt zero-based indexing. The $N$-point normalized DFT matrix denoted by $\mathbf{W}_{N}$ is defined as that the $(k,n)$th entry of $\mathbf{W}_{N}$ is $e^{-j2\pi kn/N} / \sqrt{N}$. The $i$th entry of $\mathbf{x}$ and the $(i,j)$th entry of $\mathbf{X}$ are denoted by $[\mathbf{x}]_{i}$ and $[\mathbf{X}]_{i,j}$, respectively. Given any positive integer $N$,  $\mathcal{Z}_{N}$ stands for the set $\left\{ 0,1,\cdots,N-1 \right\}$.
The root mean square value of $[\mathbf{x}]_{n}$ for all $n\in \mathcal{Z}_{N}$ is represented by $\mathrm{rms}( [\mathbf{x}]_{n} )=\sqrt{ \frac{1}{N} \left \| \mathbf{x} \right \|_{2}^{2} }$.

%==============================================================%
\section{System Model}
\label{Sec:SystemMetric}
%==============================================================%

%==============================================================%
A CPS-OFDM baseband transceiver system with the use of constellation shaping is schematized in Fig. \ref{TxRx_ConstelShapedCPSOFDM}. At the transmitter, the input $S\times 1$ data vector of the $b$th block transmission, denoted by $\mathbf{d}[b]$, is composed of $D$ quadrature amplitude modulation (QAM) symbols and $Z$ zero symbols, $S=D+Z$. For CM reduction, the data vector $\mathbf{d}[b]$ can be first shaped into an $S\times 1$ offset data vector $\mathbf{c}[b]=\mathbf{d}[b]+\mathbf{e}[b]$, where $\mathbf{e}[b]$ is a complex-valued vector introducing offset of QAM constellation points. The offset data vector $\mathbf{c}[b]$ is then fed into an $S\times S$ CPS precoding matrix defined as \cite{Huang2018_HYM}
\begin{eqnarray}
\left[ \mathbf{P} \right]_{i,kM+m} = \left[ \mathbf{p} \right]_{ {\left \langle i-kM \right \rangle _{S}} } e^{-j2\pi \frac{m}{M}i},
\end{eqnarray}
where $\mathbf{p}$ is an $S\times 1$ prototype vector producing all entries of $\mathbf{P}$, $i \in \mathcal{Z}_{S}$, $k \in \mathcal{Z}_{K}$, $m \in \mathcal{Z}_{M}$, $S=KM$. 
The prototype vector $\mathbf{p}$ can be determined by \cite[Algorithm 1]{Huang2018_HYM}.
The precoded symbols in $\mathbf{s}[b]=\mathbf{P}\mathbf{c}[b]$ are allocated to $S$ contiguous OFDM subcarriers, whose indices are given in $\mathcal{I}=\left\{ \eta, \eta+1, \cdots, \eta+S-1 \right\}$ with a starting index $\eta \geq 0$. An $N$-point inverse fast Fourier transform (IFFT) realized by $\mathbf{W}_{N}^{H}$ is used for OFDM modulation, $N \geq \eta+S$. A guard interval (GI) of length $G$ is added on each modulated signal $\mathbf{x}_{N}[b] =\left[ \mathbf{W}_{N}^{H} \right]_{\mathcal{I}}\mathbf{s}[b]$ to avoid inter-block interference (IBI) caused by multipath propagation. The GI can be chosen as either cyclic prefix (CP) or zero padding (ZP) \cite{Lin2011_TB}. Here CP insertion is considered, which can be represented by 
$\mathbf{G}=\begin{bmatrix}
\mathbf{0}_{G\times (N-G)}\quad \mathbf{I}_{G} \\
\mathbf{I}_{N}
\end{bmatrix}$.
Note that the GI utilization of the CPS-OFDM system may be optional (i.e., $\mathbf{G}=\mathbf{I}_{N}$) in some cases if IBI attributed to small tail power of each block is tolerable or negligible \cite{Huang2018_HYM}. Let $\mathbf{\Phi}=\mathbf{G} \left[ \mathbf{W}_{N}^{H} \right]_{\mathcal{I}}\mathbf{P}$ denote a synthesis matrix describing the linear structure of the transmitter. The $b$th transmitted block signal containing $N'=N+G$ samples is thus expressed as
\begin{eqnarray}
\label{TxSignal}
\mathbf{x}[b] = \mathbf{G} \mathbf{x}_{N}[b] = \mathbf{\Phi} \mathbf{c}[b] = \mathbf{G} \left[ \mathbf{W}_{N}^{H} \right]_{\mathcal{I}}\mathbf{P}\mathbf{c}[b].
\end{eqnarray}
After parallel-to-serial conversion (P/S) of (\ref{TxSignal}), the discrete-time baseband signal
\begin{eqnarray}
\label{TxSeqSignal}
{x}[n]=\sum_{b=-\infty}^{\infty} \sum_{i=0}^{S-1} \left[ \mathbf{\Phi} \right]_{\left \langle n-bN' \right \rangle _{N'},i} \left[ \mathbf{c}[b] \right]_{i}
\end{eqnarray}
is sent over a frequency-selective channel modeled as $H \left( e^{j\omega}\right)=\sum_{l=0}^{L}h[l] e^{-j\omega l} $, where $L$ and $h[l]$ stand for channel order and channel impulse response, respectively.
 
At the CPS-OFDM receiver, serial-to-parallel conversion (S/P) of ${y}[n]= \sum_{l=0}^{L }h[l]  {x}[n-l]+{z}[n]$ is first performed, where ${z}[n]$ is a complex additive white Gaussian noise (AWGN) with variance $N_{0}$. Then, the $b$th received block signal containing $N'$ samples denoted by $\mathbf{y}[b]$ is acquired. As long as $G\geq L$, the IBI component resided in $\mathbf{y}[b]$ can be eliminated by CP removal, i.e., $\mathbf{\bar G} = \left[ \mathbf{0}_{N\times G}\quad \mathbf{I}_{N} \right]$. An $N$-point FFT realized by $\mathbf{W}_{N}^{}$ is used for OFDM demodulation. Finally, the demodulated signal to be equalized can be derived as \cite{Lin2011_TB}
\begin{eqnarray}
\label{procRxSignal}
\mathbf{r}[b] =  \mathbf{H} \mathbf{P}\mathbf{c}[b]+\mathbf{z}[b] = \mathbf{H} \left[ \mathbf{P} \right]_{\mathcal{D}} \left[\mathbf{c}[b] \right]_{\mathcal{D}}+\mathbf{z}[b],
\end{eqnarray}
where $\mathcal{D}\subseteq \mathcal{Z}_{S}$ is a set of indices indicating the positions of $D$ data symbols in $\mathbf{d}[b]$, $\mathbf{H}$ is an $S\times S$ diagonal matrix with diagonal elements corresponding to channel frequency response on the occupied $S$ subcarriers, and $\mathbf{z}[b]$ is the noise vector with covariance matrix $N_{0}\mathbf{I}_{S}$. As all data symbols are zero-mean, independent and identically distributed (i.i.d.) with symbol power $E_\mathrm{s}$, the $D\times 1$ received data vector can be linearly obtained by $\mathbf{\hat d}[b] = \mathbf{Q}\mathbf{r}[b]$, where
\begin{eqnarray}
\label{MMSEFDEMx}
\mathbf{Q} = \left[ \left( \mathbf{H}   \left[ \mathbf{P} \right]_{\mathcal{D}} \right)^{H}\left( \mathbf{H}   \left[ \mathbf{P} \right]_{\mathcal{D}} \right)+\frac{N_{0}}{E_{\mathrm{s}}} \mathbf{I}_{D}\right]^{-1}\left( \mathbf{H}   \left[ \mathbf{P} \right]_{\mathcal{D}} \right)^{H},
\end{eqnarray}
is a $D\times S$ minimum mean square error (MMSE) frequency domain equalization (FDE) matrix \cite{Lin2011_TB}. It is worthy to note that the constellation shaping technique applied to the transmitter is transparent to the receiver. The locations of zero symbols of $\mathbf{c}[b]$ and $\mathbf{e}[b]$ are the same as those of $\mathbf{d}[b]$.

\subsection{Cubic Metric (CM)}
Extensive studies have shown that CM is an accurate measure of envelope fluctuation to predict the required input backoff (IBO) of the PA at the transmitter \cite{Motorola040642_3GPPTDOC,Motorola060023_3GPPTDOC,Behravan2006_WDIS_PAE,Ni2017_WDIS_PAE_CM,Zhu2014_WDIS_PAE_CM,Rahmatallah2013_WDIS_PAE_PAPR}.
The CM of the $b$th transmitted block signal \eqref{TxSignal} is defined as \cite{Motorola040642_3GPPTDOC,Motorola060023_3GPPTDOC,Behravan2006_WDIS_PAE,Ni2017_WDIS_PAE_CM,Zhu2014_WDIS_PAE_CM,Rahmatallah2013_WDIS_PAE_PAPR}
\begin{eqnarray}
\label{DefCM}
\mathrm{CM}[b] |_{\textrm{dB}} = \frac{20\log_{10} \left\{ \mathrm{RCM}[b] \right\} - \mathrm{RCM}_{\mathrm{ref}} |_{\textrm{dB}} }{ E |_{\textrm{dB}} },
\end{eqnarray}
where
\begin{eqnarray}
\label{DefRCM}
\mathrm{RCM}[b]=\mathrm{rms}\left ( \left( \frac{ \left| \left[ \mathbf{x}[b] \right]_{n} \right|}{ \mathrm{rms}\left( \left[ \mathbf{x}[b] \right]_{n} \right) } \right)^{3} \right) = N' \left( \frac{ \left \| \mathbf{x}[b]  \right \|_{6} }{ \left \| \mathbf{x}[b]  \right \|_{2} } \right)^{3}
\end{eqnarray}
is the raw CM (RCM) of \eqref{TxSignal}, $\mathrm{RCM}_{\mathrm{ref}} |_{\textrm{dB}}$ is the RCM of a reference signal in decibel, and $E |_{\textrm{dB}}$ is an empirical factor. Since $\mathrm{RCM}_{\mathrm{ref}} |_{\textrm{dB}}$ and $E |_{\textrm{dB}}$ are two constants given by the system in real cases \cite{Motorola060023_3GPPTDOC}, we are able to describe the CM \eqref{DefCM} in terms of the RCM \eqref{DefRCM} for simplicity.

\subsection{Block-wise Error Vector Magnitude (EVM)}
The key idea of constellation shaping is to impose offset on input data symbols so that the CM of each block transmission can be substantially reduced \cite{Rahmatallah2013_WDIS_PAE_PAPR,Zhu2014_WDIS_PAE_CM,Aggarwal2006_WDIS_PAE_PAPR}. In this way, there exists a difference between the actual transmitted signal \eqref{TxSignal} and the original transmitted signal written as 
\begin{eqnarray}
\label{OriginalTxSignal}
\mathbf{\tilde x}[b] = \mathbf{\Phi} \mathbf{d}[b] = \mathbf{G} \left[ \mathbf{W}_{N}^{H} \right]_{\mathcal{I}}\mathbf{P}\mathbf{d}[b].
\end{eqnarray}
To quantify the signal difference, we define the block-wise error vector magnitude (EVM) as
\begin{eqnarray}
\label{DefEVM}
\mathrm{EVM}[b] = \sqrt{ \frac{ \left \| \mathbf{x}_{}[b]  - \mathbf{\tilde x}[b]  \right \|_{2}^{2} }{  \left \| \mathbf{\tilde x}[b]  \right \|_{2}^{2} } } \leq \mathrm{EVM}_{\max},
\end{eqnarray}
where $\mathrm{EVM}_{\max}$ is the maximum allowable EVM at the transmitter. The choice of $\mathrm{EVM}_{\max}$ is relevant to tolerable detection performance degradation at the receiver. 

%As $\mathrm{EVM}_{\max}$ increases, more compromise on detection performance at the receiver would be needed.

%Different from the traditional definition of EVM, we handle 

\subsection{Out-of-Subband Emission Energy (OSBEE)}
The amount of instantaneous spectral sidelobe leakage, referred to as OSBE energy (OSBEE), is obtained by calculating the energy spectral density (ESD) of each block transmission. Specifically, the ESD of the transmitted signal \eqref{TxSignal} is
\begin{eqnarray}
\label{ESD}
\left| X_{b}\left( e^{j\omega } \right) \right|^{2} = \left| \sum_{n=0}^{N'-1} \left[ \mathbf{x}[b] \right]_{n} e^{-j\omega n} \right|^{2}, \enskip \omega \in [ -\pi,\pi ),
\end{eqnarray}
and its OSBEE is expressed as
\begin{eqnarray}
\label{OSBEE}
\mathrm{OSBEE}[b] = \int_{\omega\in \mathcal{F}_{\mathrm{OSB}}}  \left| X_{b}\left( e^{j\omega } \right) \right|^{2}   \frac{d \omega}{2\pi},
\end{eqnarray}
where $\mathcal{F}_{\mathrm{OSB}}\subset [ -\pi,\pi )$ denotes an OSB region. 
Similarly, the ESD and the OSBEE of the original transmitted signal \eqref{OriginalTxSignal} can be derived as $| \tilde{X}_{b}\left( e^{j\omega } \right) |^{2}$ and $\mathrm{OSBEE}_{\mathrm{\mathrm{ref}}}[b]$, respectively.

\subsection{Offset Vector Design Problem}
\label{subSec:Problem}
The problem of interest in this paper is to design the offset vector $\mathbf{e}[b]$ such that the RCM \eqref{DefRCM} can be reduced for each CPS-OFDM block transmission. At the same time, the EVM \eqref{DefEVM} must be constrained so as to warrant satisfactory detection reliability at the receiver. In addition, the OSBEE \eqref{OSBEE} shall also be limited in order to preserve the property of low OSBE of CPS-OFDM. The resulting mitigation of spectral regrowth and increase of spectral efficiency are expected. Since $\mathbf{e}[b]=\mathbf{c}[b]-\mathbf{d}[b]$ and $\mathbf{d}[b]$ is known at the transmitter, we will directly deal with the offset data vector $\mathbf{c}[b]$ in the next section.

%==============================================================%
\section{Proposed Method}
\label{Sec:ProposedMethod}
%==============================================================%
To tackle the problem stated in Section \ref{subSec:Problem}, we propose an offset data vector optimization method for CPS-OFDM in this section. For notational convenience, the block index ``$[b]$'' is omitted in subsequent contents. Recall that there are some zeros resided in the data vector $\mathbf{d}$ without being affected by offset values. We can only focus on $[ \mathbf{d}]_{\mathcal{D}}$, $[ \mathbf{c}]_{\mathcal{D}}$, and $[ \mathbf{\Phi}]_{\mathcal{D}}$, which will be respectively denoted by $\mathbf{\bar d}$, $\mathbf{\bar c}$, and $\mathbf{\bar \Phi}$ for brevity.

\begin{figure*}[t]
\centering \centerline{
\includegraphics[width=1.2\textwidth,clip]{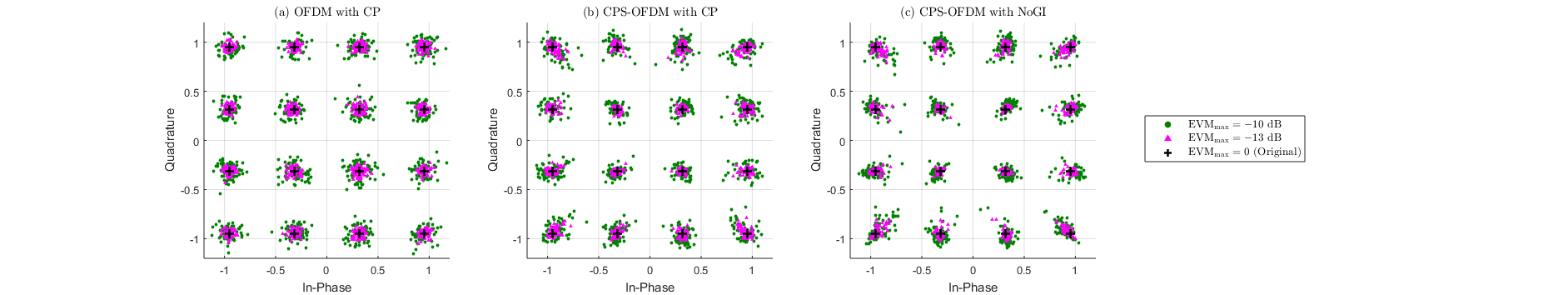}}
\caption{Scatterplots of the optimized offset data symbols in terms of different $\mathrm{EVM}_{\max}$ values for OFDM and CPS-OFDM.}
\label{Result_ConstelShapSym}
\end{figure*}

The proposed optimal offset data vector design problem is formulated as
\begin{subequations}
\label{OptOffsetVecProblem}
\begin{align}
\underset{\mathbf{\bar c} \in \mathbb{C}^{D\times 1}}{\text{minimize}} \enskip 
&\ \label{ObjFun_RCM_cvec}
\mathrm{RCM}  \\
\text{subject to} \enskip 
&\  \label{Const_EVM_cvec}
\mathrm{EVM} \leq \mathrm{EVM}_{\max}  \\
&\  \label{Const_OSBEE_cvec}
\mathrm{OSBEE} \leq \mathrm{OSBEE}_{\mathrm{ref}}.   
\end{align}
\end{subequations}
The objective \eqref{ObjFun_RCM_cvec} is to minimize the RCM of baseband CPS-OFDM signals so as to enhance PA efficiency.
However, the RCM function \eqref{DefRCM} proportional to ${ \left \| \mathbf{\bar \Phi}\mathbf{\bar c}  \right \|_{6} }/{ \left \| \mathbf{\bar \Phi}\mathbf{\bar c}  \right \|_{2} }$ makes the problem difficult to be analyzed. Hence, we treat ${ \left \| \mathbf{\bar \Phi}\mathbf{\bar c}  \right \|_{6} }$ as an alternative objective function with an additional constraint
\begin{eqnarray}
\label{TxEnergyConstraint}
{ \left \| \mathbf{\bar \Phi}\mathbf{\bar c}  \right \|_{2}^{2} } \geq { \left \| \mathbf{\bar \Phi}\mathbf{\bar d}  \right \|_{2}^{2} }
\end{eqnarray}
to ensure the resulting RCM less than the original RCM given by $N' \left( { \left \| \mathbf{\bar \Phi}\mathbf{\bar d}  \right \|_{6} }/ { \left \| \mathbf{\bar \Phi}\mathbf{\bar d}  \right \|_{2} } \right)^{3}$.
Since the constraint \eqref{TxEnergyConstraint} is nonconvex, we seek the relaxation that replaces \eqref{TxEnergyConstraint} with a looser, but convex, constraint.
According to the following formula $\left\| \mathbf{\bar \Phi}\mathbf{\bar c}- \mathbf{\bar \Phi}\mathbf{\bar d} \right\|_{2}^{2}=\left\| \mathbf{\bar \Phi}\mathbf{\bar c} \right\|_{2}^{2} - 2\Re \left\{ \mathbf{\bar c}^{H}\mathbf{\bar \Phi}^{H} \mathbf{\bar \Phi}^{} \mathbf{\bar d}^{}  \right\} + \left\| \mathbf{\bar \Phi}\mathbf{\bar d} \right\|_{2}^{2}$ \cite[Ch.~6]{Friedberg2003_TB}, the equation \eqref{TxEnergyConstraint} is equivalent to
\begin{eqnarray}
\label{normieq}
\Re \left\{ \mathbf{\bar c}^{H}\mathbf{\bar \Phi}^{H} \mathbf{\bar \Phi}^{} \mathbf{\bar d}^{}  \right\} \geq 
\left \| \mathbf{\bar \Phi}\mathbf{\bar d}  \right \|_{2}^{2}
- \frac{\left\| \mathbf{\bar \Phi}\mathbf{\bar c}- \mathbf{\bar \Phi}\mathbf{\bar d} \right\|_{2}^{2}}{2}.
\end{eqnarray}
By rewriting the EVM constraint \eqref{Const_EVM_cvec} as 
\begin{eqnarray}
\label{EVMConstraint}
{ \left \|  \mathbf{\bar \Phi}\mathbf{\bar c}  -  \mathbf{\bar \Phi}\mathbf{\bar d}  \right \|_{2}^{} } \leq  \mathrm{EVM}_{\max} {  \left \|  \mathbf{\bar \Phi}\mathbf{\bar d}  \right \|_{2}^{} },
\end{eqnarray}
it is obvious that (\ref{normieq}) can be relaxed to the convex inequality
\begin{eqnarray}
\label{cvxieq}
\Re \left\{ \mathbf{\bar c}^{H}\mathbf{\bar \Phi}^{H} \mathbf{\bar \Phi}^{} \mathbf{\bar d}^{}  \right\}  \geq 
\left \| \mathbf{\bar \Phi}\mathbf{\bar d}  \right \|_{2}^{2} \left( 1 - \frac{ \mathrm{EVM}_{\max}^{2} }{2} \right),
\end{eqnarray}
which enlarges the search space of the complex variable $\mathbf{\bar c}$ in comparison with (\ref{normieq}). 
On the other hand, the constraint \eqref{Const_OSBEE_cvec} indicates that the OSBEE of sending $\mathbf{\bar c}$ does not exceed the original OSBEE of sending $\mathbf{\bar d}$ for preserving the low-OSBE property.
In Appendix \ref{AppndixOSBEE}, we derive that the OSBEE of CPS-OFDM transmission can be expressed as a quadratic form, i.e., $\mathrm{OSBEE}=\mathbf{\bar c}^{H} \mathbf{\bar E} \mathbf{\bar c}^{}$ and $\mathrm{OSBEE}_{\mathrm{ref}}=\mathbf{\bar d}^{H} \mathbf{\bar E} \mathbf{\bar d}^{}$, where $\mathbf{\bar E}$ is a $D\times D$ Hermitian positive definite matrix. 
Therefore, the proposed optimization problem \eqref{OptOffsetVecProblem} is reformulated as
\begin{subequations}
\label{ReformOptOffsetVecProblem}
\begin{align}
\underset{\mathbf{\bar c} \in \mathbb{C}^{D\times 1}}{\text{minimize}} \enskip 
&\ \label{ObjFun_6norm_cvec}
\left \| \mathbf{\bar \Phi}\mathbf{\bar c}  \right \|_{6}  \\
\text{subject to} \enskip 
&\ \label{Const_ReFun_cvec}
\Re \left\{ \mathbf{\bar c}^{H}\mathbf{\bar \Phi}^{H} \mathbf{\bar \Phi}^{} \mathbf{\bar d}^{}  \right\}  \geq 
\left \| \mathbf{\bar \Phi}\mathbf{\bar d}  \right \|_{2}^{2} \left( 1 - \frac{ \mathrm{EVM}_{\max}^{2} }{2} \right)  \\
&\ \label{Const_2norm_cvec}
\left \|  \mathbf{\bar \Phi}\mathbf{\bar c}  -  \mathbf{\bar \Phi}\mathbf{\bar d}  \right \|_{2}^{}  \leq  \mathrm{EVM}_{\max}   \left \|  \mathbf{\bar \Phi}\mathbf{\bar d}  \right \|_{2}^{} \\
&\ \label{Const_quadraticform_cvec}
\enskip \mathbf{\bar c}^{H} \mathbf{\bar E} \mathbf{\bar c}^{} \leq \mathbf{\bar d}^{H} \mathbf{\bar E} \mathbf{\bar d}^{}.   
\end{align}
\end{subequations}
This is a convex optimization problem, of which the globally optimal point $\mathbf{\bar c}_{\mathrm{opt}}$ is guaranteed to be attained via interior-point methods \cite{Boyd2004_TB}.
To solve the problem \eqref{ReformOptOffsetVecProblem}, we adopt \texttt{CVX}, a package for specifying and solving convex programs \cite{CVXtool}.

%==============================================================%

%==============================================================%
\section{Simulation Results}
\label{Sec:SimResult}
%==============================================================%
In this section, the performance evaluations of the proposed method are provided in terms of RCM, OSBE, bit-error rate (BER), and spectral efficiency with practical parameters.

\subsection{Simulation Parameters and Assumptions}
Referring to \cite[Annex A.1.1]{3GPPTR38802}, the system parameters are chosen as follows.
The carrier frequency is at $4$ GHz. 
The sampling rate is $1.92$ MHz.
The FFT size $N$ is $128$ with $15$ kHz subcarrier spacing.
The CP length $G$ is $9$ (or $G=0$ without GI (NoGI)).
For the settings of CP and NoGI, a transmission time interval (TTI) of $1$ ms can accommodate $14$ and $15$ blocks, respectively.
Uncoded 16-QAM is utilized with ${E_{\mathrm{s}}}=1$. 
The bit power versus noise variance is shown as ${E_{\mathrm{b}}}/N_{0}$.
The bandwidth assigned to a target user, denoted by $\mathrm{BW}$, is $720$ kHz, namely, $S=48$, $ \mathcal{I}_{}=\left\{ 28,29,\cdots, 75 \right\}$.
The ninth-order polynomial approximation specified in \cite{3GPP166004_3GPPTDOC} is used to model PA nonlinearity with phase compensation of $76.3$ degrees \cite{Huawei166093_3GPPTDOC}. 
Considering that a low-cost machine transmitter often desires sufficiently high power efficiency even with some nonlinear signal distortion \cite{Behravan2009_WDIS_PAE_VIP}, we select a quite challenging IBO value of $0$ dB (i.e., operating at the saturation point of the PA) to investigate the achievable performance.
The spectrum emission mask (SEM) defined in \cite[Table 6.6.2.1.1-1]{3GPPTS36101} serves as a reference for the required size of guard band $\Delta$, while the spectral regrowth due to PA nonlinearity is with the maximum transmit power of $22$ dBm.
The configuration of single transmit antenna and single receive antenna (1T1R) is set. 
The tapped delay line (TDL)-C channel model is used in terms of the delay spread scaled by $300$ ns with $3$ km/h mobility \cite[Table 7.7.2-3]{3GPPTR38901}.
The MMSE-FDE (\ref{MMSEFDEMx}) is adopted at the receiver under the assumptions of perfect synchronization and channel estimation.

CPS-OFDM is parameterized by $K=2$, $M=24$, $D=46$, $Z=2$, $\mathcal{D}=\{ 1,\cdots,23, 25,\cdots,47 \}$, and the prototype vector $\mathbf{p}$ determined by \cite[Algorithm 1]{Huang2018_HYM} (with the same factors such as $\beta=10$ and $\epsilon=0$ specified in \cite[Sec. V-C]{Huang2018_HYM}). The OSB range $\mathcal{F}_{\mathrm{OSB}}$ to be concerned corresponds to the subcarriers indexed by $\{ 0,\cdots,23, 80,\cdots,127 \}$.
The settings of CP and NoGI in the CPS-OFDM system are taken into account. 
The results of adopting traditional OFDM ($\mathbf{P}=\mathbf{I}_{48}$) with CP and $Z=0$ serve as performance benchmarks. The optimal offset data vector $\mathbf{\bar c}_{\mathrm{opt}}$ can be found by solving the proposed constellation shaping optimization problem \eqref{ReformOptOffsetVecProblem}, in which $\mathrm{EVM}_{\max}$ chosen as $-13$ dB and $-10$ dB are of interest.

\begin{figure}[t]
\centering \centerline{
\includegraphics[width=0.5\textwidth,clip]{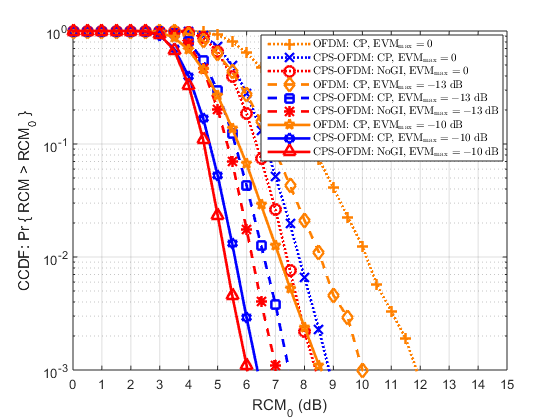}}
\caption{The proposed optimal offset data vector design \eqref{ReformOptOffsetVecProblem} can further reduce the RCM of CPS-OFDM signals to meet the requirement of high PA efficiency.}
\label{Result_CM}

\centering \centerline{
\includegraphics[width=0.5\textwidth,clip]{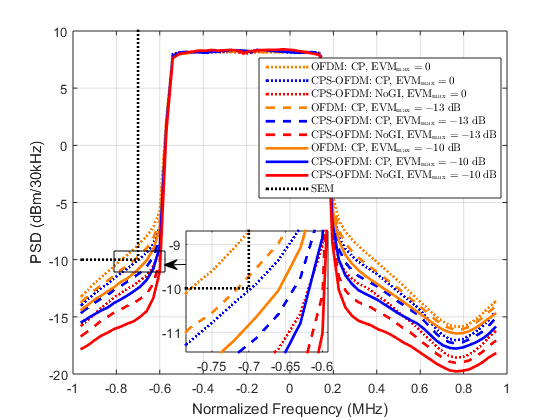}}
\caption{Simulated PSD results ($\mathrm{IBO}=0$ dB) that the spectral containment of CPS-OFDM can be further improved by the proposed scheme \eqref{ReformOptOffsetVecProblem}.}
\label{Result_PSDPA_IBO0}
\end{figure}

\begin{figure}[t]
\centering \centerline{
\includegraphics[width=0.5\textwidth,clip]{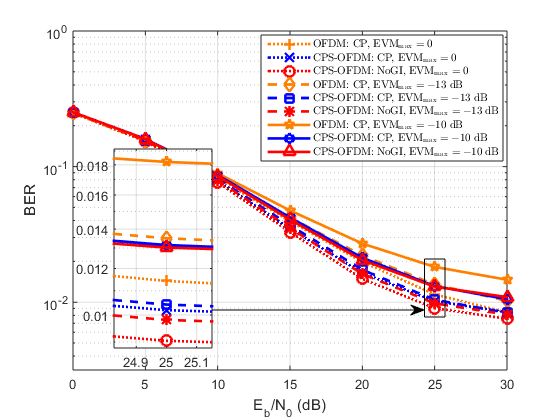}}
\caption{Uncoded 16-QAM BER results that the utilization of constellation shaping is accompanied with slight detection performance degradation.}
\label{Result_BER_IBO0_MMSEPA_TDLC300}

\centering \centerline{
\includegraphics[width=0.5\textwidth,clip]{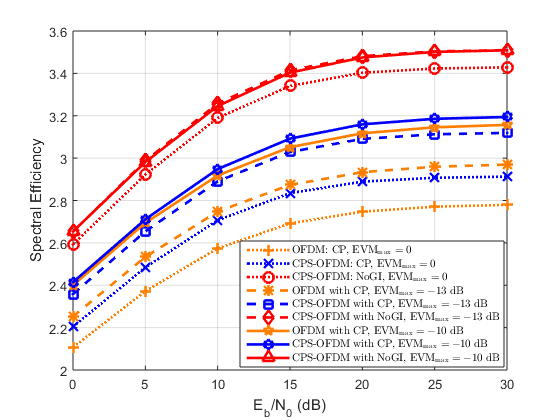}}
\caption{Spectral efficiency improvement through the proposed constellation shaping technique mainly due to the decrease of the required guard bands.}
\label{Result_SE_IBO0_MMSEPA_TDLC300}
\end{figure}

\subsection{Performance Evaluations}
The scatterplots of the optimized offset data symbols in $\mathbf{\bar c}_{\mathrm{opt}}$ in terms of different $\mathrm{EVM}_{\max}$ values for OFDM, CPS-OFDM with CP, and CPS-OFDM with NoGI are illustrated in Fig. \ref{Result_ConstelShapSym}.
As $\mathrm{EVM}_{\max}$ increases, more offset imposed on the original constellation points can be observed. 

CM performance is assessed by the empirical complementary cumulative distribution function (CCDF) curve of RCM outcomes.
The CCDF is defined as $1-\mathrm{Pr}\left\{ \mathrm{RCM} \leq \mathrm{RCM}_{0} \right\}$, where $\mathrm{Pr}\left\{ \mathrm{RCM} \leq \mathrm{RCM}_{0} \right\}$ means the probability of a RCM value that does not exceed a given threshold $\mathrm{RCM}_{0}$. 
In Fig. \ref{Result_CM}, it is shown that the proposed optimal offset data vector design \eqref{ReformOptOffsetVecProblem} can further reduce the RCM of CPS-OFDM signals to meet the requirement of rather high PA efficiency.
Note that the setting of NoGI is basically more friendly to the PA than the setting of CP at the CPS-OFDM transmitter \cite{Huang2018_HYM}. 

Owing to strong PA nonlinearity usually causing severe spectral regrowth \cite{Cottais2008_WDIS_SR}, the resulting OSBE without IBO must be carefully treated. 
To present OSBE performance, we simulate the power spectral density (PSD) of \eqref{TxSeqSignal} by $\frac{1}{B}\sum_{b=0}^{B-1} \left| X_{b}\left( e^{j\omega } \right) \right|^{2} $, where $B=10^{4}$ is the number of transmitted blocks.
As revealed in Fig. \ref{Result_PSDPA_IBO0}, RCM reduction (see Fig. \ref{Result_CM}) is also much helpful to spectral regrowth mitigation.
Using the proposed scheme, the spectral containment property of CPS-OFDM can be further improved. 

The BER results are given in Fig. \ref{Result_BER_IBO0_MMSEPA_TDLC300}.
It can be found that the use of constellation shaping leads to slight performance degradation. 
Although there exists IBI, here CPS-OFDM with NoGI possesses the best detection reliability mainly because of the least PA nonlinear distortion as compared to the others. 

The spectral efficiency (SE) is calculated by $\mathrm{SE} = \frac{N_{\mathrm{bit}}DN_{\mathrm{B}}(1-\mathrm{BER})}{\mathrm{TTI} \cdot (\mathrm{BW}+\Delta)} $ \cite{3GPPTR38802}, where $N_{\mathrm{bit}}=4$ is the number of bits per 16-QAM data symbol, $N_{\mathrm{B}}=14$ (or 15 for NoGI) is the number of transmitted blocks per TTI, and $\mathrm{BER}$ can be obtained from the results shown in Fig. \ref{Result_BER_IBO0_MMSEPA_TDLC300}.
The required guard band $\Delta$ is decided such that the OSBE can be lower than the SEM requirement of $-10$ dBm per $30$ kHz measurement bandwidth (see Fig. \ref{Result_PSDPA_IBO0}). 
The SE comparisons of different waveform settings are summarized in Fig. \ref{Result_SE_IBO0_MMSEPA_TDLC300}.
The proposed constellation shaping optimization method is validated to be able to further increase the SE of CPS-OFDM.

%==============================================================%
\section{Conclusion}
\label{Sec:Conclusion}
%==============================================================%
A constellation shaping optimization method is proposed to reduce the cubic metric (CM) of CPS-OFDM signals. 
The optimized offset data symbols also preserve the original low-OSBE property of CPS-OFDM.
Benefited from the extremely low CM, the spectral regrowth and the signal distortion caused by severe power amplifier (PA) nonlinearity can be significantly alleviated.
Therefore, the resulting gain in spectral efficiency can be found.
The future work is to develop a customized interior-point method for the proposed optimization problem in order to ease the real-time computational burden.

%==============================================================%
\appendices
%\section{}
\section{Derivation of OSBEE in Quadratic Form}
\label{AppndixOSBEE}
Recall that the transmitted block signal \eqref{TxSignal} can be expressed as $\mathbf{x}=\mathbf{\Phi}\mathbf{c}$, where the block index ``$[b]$'' has been omitted.
The $n$th element of $\mathbf{x}$ is written as
\begin{eqnarray}
\label{TxSingalElement}
x_{n} = \sum_{i=0}^{S-1} \left[ \mathbf{\Phi} \right]_{n,i} \left[ \mathbf{c} \right]_{i}.
\end{eqnarray}
Using $i=kM+m$, the equation \eqref{TxSingalElement} is rewritten as 
\begin{eqnarray}
\label{TxSingalElementExpand}
x_{n} = \sum_{k=0}^{K-1} \sum_{m=0}^{M-1} \left[ \mathbf{\Phi} \right]_{n,kM+m} \left[ \mathbf{c} \right]_{kM+m}, 
\end{eqnarray}
where we can expand each entry of the synthesis matrix $\mathbf{\Phi}$ as 
\begin{align}
\left[ \mathbf{\Phi} \right]_{n,kM+m}  =& \nonumber \\
\frac{1}{\sqrt{NM}}  & \sum_{i=0}^{S-1} \left[ \mathbf{p} \right]_{ {\left \langle i-kM \right \rangle _{S}} }  e^{-j \frac{2\pi}{M}im} e^{j \frac{2\pi}{N}(\eta+i)(n-G)}.  \nonumber
\end{align}
By performing $X(e^{j\omega})=\sum_{n=0}^{N+G-1} x_{n} e^{-j\omega n}$, the magnitude response of \eqref{TxSingalElementExpand} is given by
\begin{align}
\label{MagX}
\left| X(e^{j\omega}) \right| & = \frac{1}{\sqrt{NM}} 
\bigg| \sum_{k=0}^{K-1} \sum_{m=0}^{M-1}  \left[ \mathbf{c} \right]_{kM+m} \cdot \nonumber \\
\sum_{i=0}^{S-1} & \left[ \mathbf{C}_{kM}\mathbf{p} \right]_{i}  e^{-j2\pi i \left( \frac{m}{M} + \frac{G}{N} \right)} W\left( e^{j \left[ \omega- \frac{2\pi}{N} (\eta + i) \right]} \right)
\bigg|,
\end{align}
where $\mathbf{C}_{kM}$ is a downshift permutation matrix defined as 
\[
\mathbf{C}_{0}=\mathbf{I}_{S}, \enskip
\mathbf{C}_{kM}=\begin{bmatrix}
\mathbf{0} & \mathbf{I}_{kM} \\
\mathbf{I}_{S-kM} & \mathbf{0}
\end{bmatrix},
\]
and $W\left( e^{j \left[  \omega-\frac{2\pi}{N}(\eta+i) \right]} \right)=\sum_{n=0}^{N+G-1} e^{-j \left[ \omega - \frac{2\pi}{N} (\eta +i)\right]n} $.
We define an $S\times 1$ vector $\mathbf{w}_{m}\left( e^{j\omega} \right)$ whose the $i$th entry is
\begin{align}
\left[ \mathbf{w}_{m}\left( e^{j\omega } \right) \right]_{i} =\frac{1}{\sqrt{NM}}e^{j 2\pi i \left( \frac{m}{M}+\frac{G}{N} \right) }W^{*}\left( e^{j \left[  \omega-\frac{2\pi}{N}(\eta+i) \right]} \right). \nonumber
\end{align}
Then, the equation \eqref{MagX} can be simplified as 
\begin{eqnarray}
\left| X(e^{j\omega}) \right|  &=& \left| \sum_{k=0}^{K-1} \sum_{m=0}^{M-1} 
\left[ \mathbf{w}_{m}^{H} \left( e^{j\omega } \right)  \mathbf{C}_{kM}\mathbf{p} \right] \cdot
\left[ \mathbf{c} \right]_{kM+m} \right| \nonumber \\ 
&=& \left| \mathbf{u}^{H} \left( e^{j\omega} \right) \mathbf{c} \right|
\end{eqnarray}
where the $(kM+m)$th entry of the $S\times 1$ vector $\mathbf{u} \left( e^{j\omega} \right)$ is
\[
\left[ \mathbf{u} \left( e^{j\omega} \right) \right]_{kM+m} = 
\left[ \mathbf{w}_{m}^{H} \left( e^{j\omega } \right)  \mathbf{C}_{kM}\mathbf{p} \right]^{H}.
\] 
Let $ \mathbf{U}^{} \left( e^{j\omega} \right) = \mathbf{u}^{} \left( e^{j\omega} \right)  \mathbf{u}^{H} \left( e^{j\omega} \right)  $, the ESD of \eqref{TxSingalElementExpand} is obtained by
\begin{eqnarray}
\left| X(e^{j\omega}) \right|^{2} =   \left| \mathbf{u}^{H} \left( e^{j\omega} \right) \mathbf{c} \right|^{2}
= \mathbf{c}^{H}  \mathbf{U}^{} \left( e^{j\omega} \right)  \mathbf{c}^{}.
\end{eqnarray}
Therefore, the OSBEE of one CPS-OFDM block transmission with constellation shaping can be formulated as a quadratic form as below
\begin{eqnarray}
\mathrm{OSBEE} 
= \int_{\omega\in \mathcal{F}_{\mathrm{OSB}}} \left| X(e^{j\omega}) \right|^{2}   \frac{d \omega}{2\pi}
= \mathbf{c}^{H}  \mathbf{E}^{}  \mathbf{c}^{} 
= \mathbf{\bar c}^{H} \mathbf{\bar E} \mathbf{\bar c}^{},
\end{eqnarray}
where  $\mathbf{E} = \int_{\omega\in \mathcal{F}_{\mathrm{OSB}}}  \mathbf{U} \left( e^{j\omega } \right)  \frac{d \omega}{2\pi} $ and $\mathbf{\bar E} = \left[ \mathbf{E} \right]_{\mathcal{D},\mathcal{D}}$ are two Hermitian positive definite matrices, $\mathbf{\bar c} = [ \mathbf{c}]_{\mathcal{D}}$.
Similarly, the original OSBEE of one CPS-OFDM block transmission can be derived as $\mathrm{OSBEE}_{\mathrm{ref}}  = \mathbf{\bar d}^{H} \mathbf{\bar E} \mathbf{\bar d}^{}$.

%Let $\mathcal{K} \subseteq \mathcal{Z}_{K}$ and $\mathcal{M} \subseteq \mathcal{Z}_{M}$ be two sets of indices corresponding to the positions of $D$ offset data symbols in $\mathbf{c}$, $D= \left| \mathcal{D} \right| =\left| \mathcal{K} \right| \left| \mathcal{M} \right| \leq S$.

%\section{Proof of Lemma 1}
%\label{appendixLemma1}

%\section*{Acknowledgment}
%The authors would like to thank the anonymous reviewers for their useful suggestions, which substantially helped to improve the quality of the paper.
%==============================================================%
%======================= Bibliography =========================%
%==============================================================%

%==============================================================%
%==============================================================%
%==============================================================%

\begin{thebibliography}{99}

\bibitem{3GPPTR38913}
3GPP, ``Study on scenarios and requirements for next generation access technologies,'' Technical Report (TR) 38.913, V14.3.0, June 2017.

\bibitem{Sexton2018_WD}
C. Sexton, Q. Bodinier, A. Farhang, N. Marchetti, F. Bader, and L. A. DaSilva, ``Enabling asynchronous machine-type D2D communication using multiple waveforms in 5G,'' \emph{IEEE Internet of Things Journal}, vol. 5, no. 2, pp. 1307-1322, Apr. 2018.

\bibitem{Wunder2014_WD}
G. Wunder, P. Jung, M. Kasparick, T. Wild, F. Schaich, Y. Chen, S. T. Brink, I. Gaspar, N. Michailow,  A.  Festag, L. Mendes, N. Cassiau, D. Ktenas, M. Dryjanski, S. Pietrzyk, B. Eged, P. Vago, and F. Wiedmann, ``5GNOW: non-orthogonal, asynchronous waveforms for future mobile applications,'' \emph{IEEE Communications Magazine}, vol. 52, no.2, pp. 97-105, Feb. 2014.

\bibitem{Huang2016_HYM}
Y. Huang, B. Su, and I-K. Fu, ``Heterogeneous LTE downlink spectrum access using embedded-GFDM,'' in \emph{Proc. International Conference on Communication (ICC) Workshop on 5G RAN Design}, May 2016.


\bibitem{3GPPTR38912}
3GPP, ``Study on New Radio (NR) access technology,'' Technical Report (TR) 38.912, V14.1.0, June 2017.

\bibitem{Lien2017_HYM}
S.-Y. Lien, S.-L. Shieh, Y. Huang, B. Su, Y.-L. Hsu, and H.-Y. Wei, ``5G new radio: waveform, frame structure, multiple access, and initial access,'' \emph{IEEE Communications Magazine}, vol. 55, pp. 64-71, June 2017.

\bibitem{Ankarali2017_WD}
Z. E. Ankarali, B. Pekoz, and H. Arslan, ``Flexible radio access beyond 5G: a future projection on waveform, numerology, and frame design principles,'' \emph{IEEE Access}, vol. 5, pp. 18295-18309, Sep. 2017.

\bibitem{Zaidi2016_WD}
A. A. Zaidi, R. Baldemair, H. Tullberg, H. Bjorkegren, L. Sundstrom, J. Medbo, C. Kilinc, and I. D. Silva, ``Waveform and numerology to support 5G services and requirements,'' \emph{IEEE Communications Magazine}, vol. 54, no. 11, pp. 90-98, Nov. 2016.

\bibitem{Huang2018_HYM}
Y. Huang and B. Su, ``Circularly pulse-shaped precoding for OFDM: a new waveform and its optimization design for 5G New Radio,'' Submitted to \emph{IEEE Access Special Section on New Waveform Design and Air-Interface for Future Heterogeneous Network towards 5G}, 2018. [Online]. Available: http://arxiv.org/abs/1805.06775


\bibitem{Zhang2016_WD}
X. Zhang, L. Chen, J. Qiu, and J. Abdoli, ``On the waveform for 5G,'' \emph{IEEE Communications Magazine}, vol. 54, no. 11, pp. 74-80, Nov. 2016.

\bibitem{Motorola040642_3GPPTDOC}
Motorola, ``R1-040642: Comparison of PAR and cubic metric for power de-rating,'' in 3GPP TSG RAN WG1 Meeting \#37, Montreal, Canada, May 10-14, 2004.

\bibitem{Motorola060023_3GPPTDOC}
Motorola, ``R1-060023: Cubic metric in 3GPP-LTE,'' in 3GPP TSG RAN WG1 LTE Adhoc Meeting, Helsinki, Finland, Jan. 23-26, 2006.

\bibitem{Behravan2006_WDIS_PAE}
A. Behravan and T. Eriksson, ``Some statistical properties of multicarrier signals and related measures,'' in  \emph{Proc. IEEE 63rd Vehicular Technology Conference}, pp. 1854-1858, May 2006.

\bibitem{Ni2017_WDIS_PAE_CM}
C. Ni and T. Jiang, ``Minimizing the error vector magnitude with constrained cubic metric and spectral sidelobe in NC-OFDM-based cognitive radio systems,'' \emph{IEEE Trans. on Vehicular Technology}, vol. 66, no. 1, pp. 358-363, Jan. 2017.

\bibitem{Zhu2014_WDIS_PAE_CM}
X. Zhu, H. Hu, Z. Meng, and J. Xia, ``On minimizing the cubic metric of OFDM signals using convex optimization,'' \emph{IEEE Trans. on Broadcasting}, vol. 60, no. 3, pp. 511-523, Sep. 2014.

\bibitem{Rahmatallah2013_WDIS_PAE_PAPR}
Y. Rahmatallah and S. Mohan, ``Peak-to-average power ratio reduction in OFDM systems: a survey and taxonomy,'' \emph{IEEE Communications Surveys \& Tutorials}, vol. 15, no. 4, Fourth Quarter 2013.

\bibitem{Aggarwal2006_WDIS_PAE_PAPR}
A. Aggarwal and T. H. Meng, ``Minimizing the peak-to-average power ratio of OFDM signals using convex optimization,'' \emph{IEEE Trans. on Signal Processing}, vol. 54, no. 8, pp. 3099-3110, Aug. 2006.

\bibitem{Mattingley2010_OPT}
J. Mattingley and S. Boyd, ``Real-time convex optimization in signal processing,'' \emph{IEEE Signal Processing Magazine}, vol. 27, pp. 50-61, May 2010.

\bibitem{Lin2011_TB}
Y.-P. Lin, S.-M. Phoong, and P. P. Vaidyanathan, \emph{Filter Bank Transceivers for OFDM and DMT Systems}, Cambridge University Press, 2011.

\bibitem{Friedberg2003_TB}
S. H. Friedberg, A. J. Insel, and L. E. Spence, \emph{Linear Algebra}, 4th ed., Pearson Education Inc., 2003.

\bibitem{Boyd2004_TB}
S. Boyd and L. Vandenberghe, \emph{Convex Optimization}, Cambridge University Press, 2004.

\bibitem{CVXtool}
CVX Research, Inc. CVX: Matlab software for disciplined convex programming, version 2.1. \url{http://cvxr.com/cvx}, Dec. 2016.

\bibitem{3GPPTR38802}
3GPP, ``Study on New Radio access technology physical layer aspects,'' Technical Report (TR) 38.802, V14.2.0, Sep. 2017.

\bibitem{3GPP166004_3GPPTDOC}
3GPP, ``R1-166004: Response LS on realistic power amplifier model for NR waveform evaluation ,'' in 3GPP TSG RAN WG1 Meeting \#85, Nanjing, China, May 23-27, 2016.

\bibitem{Huawei166093_3GPPTDOC}
Huawei and HiSilicon, ``R1-166093: Waveform evaluation updates for case 1a and case 1b,'' in 3GPP TSG RAN WG1 Meeting \#86, Gothenburg, Sweden, Aug. 22-26, 2016.

\bibitem{Behravan2009_WDIS_PAE_VIP}
A. Behravan and T. Eriksson, ``Tone reservation to reduce the envelope fluctuations of multicarrier signals,'' \emph{IEEE Trans. on Wireless Communications}, vol. 8, no. 5, pp. 2417-2423, May 2009.

\bibitem{3GPPTS36101}
3GPP, ``User Equipment (UE) radio transmission and reception,'' Technical Specification (TS) 36.101, V15.1.0, Dec. 2017.

\bibitem{3GPPTR38901}
3GPP, ``Study on channel model for frequencies from 0.5 to 100 GHz ,'' Technical Report (TR) 38.901, V14.3.0, Dec. 2017.

\bibitem{Cottais2008_WDIS_SR}
E. Cottais, Y. Wang, and S. Toutain, ``Spectral regrowth analysis at the output of a memoryless power amplifier with multicarrier signals,'' \emph{IEEE Trans. on Communications}, vol. 56,  no. 7, pp. 1111-1118, July 2008.





\end{thebibliography}
\end{document}